# Evolution from Landau Quantization to Discrete Scale Invariance Revealed by Quantum Oscillations in Topological Materials

Jiayi Yang[1,2#], Nannan Tang[1#], Yunxing Li[1#], Jiawei Luo[3#], Huakun Zuo[4], Gangjian Jin[4], Ziqiao Wang[5], Haiwen Liu[6], Yanzhao Liu[5*], Donghui Guo[1], X.C. Xie[2,7,8], Jian Wang[2,7,8*], Huichao Wang[1,4*]

[1]*Guangdong Provincial Key Laboratory of Magnetoelectric Physics and Devices, Center for Neutron Science and Technology, School of Physics, Sun Yet-sen University, Guangzhou, China*

[2]*International Center for Quantum Materials, School of Physics, Peking University, Beijing, China.*

[3]*State Key Laboratory of Quantum Functional Materials & ShanghaiTech Laboratory for Topological Physics, School of Physical Science and Technology, ShanghaiTech University, Shanghai, 201210, China*

[4]*Wuhan National High Magnetic Field Center, Huazhong University of Science & Technology, Wuhan, China*

[5]*Quantum Science Center of Guangdong–Hong Kong–Macao Greater Bay Area (Guangdong), Shenzhen, China*

[6]*Center for Advanced Quantum Studies, School of Physics and Astronomy, Beijing Normal University, Beijing 100875, China*

[7]*Hefei National Laboratory, Hefei 230088, China*

[8]*Collaborative Innovation Center of Quantum Matter, Beijing, China*

[#]These authors contribute equally: Jiayi Yang, Nannan Tang, Yunxing Li, Jiawei Luo.

[*]Corresponding authors. Email: wanghch26@mail.sysu.edu.cn (H.W.); liuyanzhao@quantumsc.cn (Y.Liu); jianwangphysics@pku.edu.cn (J.W.).

**Abstract**

Dirac materials have been a unique solid-state platform for exploring relativistic quantum phenomena including supercritical atomic collapse, which leads to emergent discrete scale symmetry and log-periodic quantum oscillations. In the relativistic regime, the fundamental effect in quantum electrodynamics, vacuum polarization, can further modulate the atomic collapse-like state by screening bare charges but is rarely harnessed in condensed matter system. Here, we report a continuous progression from low-field Shubnikov–de Haas oscillations to high-field log-periodic oscillations in the Dirac material $HfTe_5$, with both phenomena modulated by Fermi surface anisotropy. This maps the transition from single-particle Landau levels to an interaction-driven, discrete scale-invariant energy spectrum of quasi-bound states. Crucially, our findings suggest vacuum polarization provides a compelling mechanism for renormalizing the effective impurity charge, quantitatively explaining the carrier-density-dependent scale factor. By revealing the intricate interplay between Landau quantization, many-body electronic screening, and scale-symmetry breaking, our results establish Dirac solids as a controllable platform for exploring relativistic vacuum effects and emergent novel symmetry.

**Introduction**

Symmetry stands as a central concept and fundamental principle in modern physics. Scale symmetry, a universal feature in complex systems, describes the self-similarity or invariance of a system at different scales[1]. When the scales are fixed to be certain geometrical series, the symmetry is broken into novel discrete scale invariance (DSI), which remains challenging to realize and detect in quantum systems[2-9]. Dirac materials have recently emerged as ideal platforms for exploring DSI[10-17]. Their low-energy quasiparticles obey massless Dirac equations, and in the presence of a Coulomb potential, the system exhibits a scale invariant Hamiltonian since both the kinetic and the interaction terms scale as $1/r$. When boundary conditions are imposed, DSI may emerge in the supercritical regime[18]. More remarkably, such systems provide a unique opportunity to explore the long-standing atomic collapse[19,20]—a fundamental prediction of quantum electrodynamics (QED) that has remained experimentally inaccessible in nuclear physics due to the inaccessible extreme condition $Z\alpha > 1$, where $Z$ is the nuclear charge and the fine structure constant $\alpha \approx 1/137$. In Dirac materials, the effective fine-structure constant ($\alpha^* = \alpha c/v_F$) is significantly enhanced because the Fermi velocity ($v_F$) is much smaller than light speed ($c$)[21,22]. The large $\alpha^*$ enables the realization of supercritical Coulomb potential and the formation of quasi-bound states analogous to atomic collapse. This system serves as a controllable and unique platform for investigating phenomena of relativistic quasi-particles.

Under strong magnetic fields, these discrete-scale-invariant quasi-bound states resonantly scatter itinerant carriers at the Fermi surface, giving rise to exotic log-periodic quantum oscillations (QOs) [11, 12]. Such QOs have been discovered in topological transition-metal pentatelluride[11, 23, 24] and Weyl semiconductor[25, 26], providing a direct signature of DSI and opening a new frontier in probing relativistic QED within solid-state systems.

The log-periodic QOs generally emerge at low temperatures and high magnetic fields[11, 23-26], conditions similar to those that produce the well-known Shubnikov–de Haas (SdH) QOs. However, the relationship between these two fundamental phenomena has remained unclear, as they have not been observed simultaneously in a single system. As known, the cyclotron motion of charge carriers perpendicular to the magnetic field is quantized into discrete Landau levels (Fig. 1a), giving rise to the well-established SdH QOs periodic in $1/B$—a key technique for mapping Fermi surfaces and band structures[27]. With increasing magnetic field, the system approaches the quantum limit (QL), where all carriers are confined to the lowest Landau level[28]. In this extreme condition, the kinetic energy is effectively quenched and the interaction becomes crucial[29-32]. Theoretically, the log-periodic QOs are expected to occur only in this regime due to supercritical Coulomb interactions (Fig. 1a). In this supercritical regime, the vacuum polarization effect, which has been extensively studied within relativistic QED, is predicted to renormalize the effective charge, and thus provides a promising avenue for modifying the DSI scaling behavior (Fig. 1a). Despite these compelling theoretical prospects, experimental evidence remains elusive. Previous experimental studies of the DSI in topological materials lacked a definitive identification of the QL via the SdH effect, leaving the transition from Landau quantization to DSI incomplete. Furthermore, while log-periodic QOs serve as a powerful probe into the fractal-like energy spectrum beyond the Landau paradigm, it remains unclear how the scale factor is quantitatively reshaped by its electronic environment. Specifically, how the Dirac sea acts as a many-body medium to renormalize scale symmetry via vacuum polarization remains a fundamental question in solid-state physics. To address these gaps, carrier-tunable samples and systematical magnetotransport measurements are required to demonstrate the full evolution from the Landau level physics to DSI in QED and establish a unified picture to link these two different regimes.

In this work, the transition from SdH to log-periodic QOs is bridged within a single material system under high magnetic fields. Utilizing high-quality *p*-type HfTe$_5$ single crystals with tunable carrier densities, we first characterize the previously elusive hole pocket through detailed angular-dependent SdH QOs at low magnetic fields, resolving its Fermi surface anisotropy, effective mass, and Landé *g*-factor. When the magnetic field is increased, clear log-periodic QOs emerge in the QL regime, marking the evolution from Landau quantization to DSI within the same samples. Interestingly, by realizing tunable log-periodic QOs in samples with different carrier densities, we uncover the notable role of vacuum polarization in defining the DSI landscape. We establish a quantitative link between the scale factor and the electronic environment, demonstrating that vacuum polarization is an important mechanism that dictates the scaling laws of relativistic quasiparticles. These results confirm that log-periodic QOs develop only in the QL regime, succeeding the SdH QOs. To our knowledge, this study presents the first experimental demonstration of a continuous evolution between SdH and log-periodic QOs. Systematically tracking the evolution of these two fundamental QOs allows us to establish a comprehensive physical framework for these two quantum regimes, and provide a quantitatively connection between them. These findings offer key insights into the electronic spectrum of Dirac materials and advance the understanding of relativistic quantum phenomena in condensed matter systems, where vacuum polarization can act as a primary tuning knob.

**Tunable hole doping in HfTe$_5$ single crystals**

A series of high-quality HfTe$_5$ single crystals were grown using the flux method (see Methods). Hall measurements (inset of Fig. 1b) of a typical sample (A11) reveal *p*-type transport from 2 to 300 K, consistent with previously reported flux-grown crystals[33, 34]. The temperature-dependent resistivity ($\rho-T$) varies across samples (Fig. 1b), showing different resistivity peak temperature ($T_\mathrm{p}$) and residual resistivity at 2 K. Figure 1c presents the magnetoresistance (*MR* = 100% × (*R*(*H*) – *R*(0)) / *R*(0)) curves at low temperatures for multiple samples. In samples with lower carrier densities (e.g. C2), the MR shows no discernible oscillations in the low magnetic fields below 1 T (Fig. S1). For samples with higher carrier densities (e.g. A1 to A10), pronounced QOs are observed in MR. Moreover, these oscillations are evenly spaced in 1/*B* (Fig. S2), confirming their origin as the SdH effect. Landau fan diagrams yield SdH oscillation frequencies $F$ ranging from 0.99 T to 1.81 T (Fig. S2). Using the Lifshitz-Onsager relation $F = (\hbar/2\pi e) \cdot S_F$ between the SdH oscillation frequency and the Fermi surface cross-section area $S_F = \pi k_F^2$, we estimate the hole density under the assumption of an isotropic Fermi surface as $n = \frac{1}{3\pi^2}(k_F)^3 = \frac{1}{3\pi^2}\left(F\frac{2e}{\hbar}\right)^{\frac{3}{2}}$, giving $n \approx$ 5.57×10$^{15}$ to 1.38×10$^{16}$ cm$^{-3}$ across the samples[35]. Although the Fermi surface of HfTe$_5$ shows anisotropy as reported[36], the estimations here demonstrate the variation in carrier density. Figure 1d summarizes the values of $n$ in different HfTe$_5$ samples, along with the different $T_p$ further confirming the tunable hole doping. For higher-density samples, to achieve the thermal excitation regime needs a higher temperatures and thus shows a larger $T_p$[37]. The series of density-tunable crystals offers a unique platform to track the evolution of both types of QOs and to understand the transition from Landau quantization to DSI under strong magnetic fields.

**SdH and log-periodic QOs in a single sample**

To investigate the evolution of QOs over different magnetic field regimes, we performed MR measurements under pulsed ultrahigh magnetic fields up to 56 T. Figure 2a presents the MR behavior of sample A4 at different temperatures. Clear QOs are observed throughout the field range. By subtracting a smooth background to remove the non-oscillatory contribution, we obtain the oscillatory component of the MR. In the low-field regime, the QOs show a periodicity in $1/B$ (Fig. 2b), with the local peaks (valleys) marked by solid (dashed) lines. The Landau index ($\nu$) analyses further confirm the periodicity in $1/B$ (Fig. 2c) rather than a log-periodic behavior (Fig. 2d). For these low-field SdH QOs, the frequency $F$ is revealed to be around 1.15 T based on the Landau fan diagram (Fig. 2c). In contrast, as shown in Fig. 2e, the MR under high magnetic fields exhibits clear QOs that become periodic when plotted on semi-logarithmic scale, with the extrema similarly indicated. Further index analyses ($i$=1 denotes the first peak beyond the SdH QOs, and so on) confirm that these QOs possess log$B$ periodicity (Fig. 2f) rather than $1/B$ periodicity (Fig. 2g). The log-periodicity signals DSI feature[38], which can be described by a characteristic scale factor $\lambda = B_{i+1}/B_i = 10^b = 2.57$ with $b$ obtained from the linear slope in Fig. 2f. These results reveal a clear transition from low-field SdH QOs to log-periodic

QOs at high fields. These results reveal a clear transition from low-field SdH QOs to log-periodic QOs at high fields within a single sample.

The clear SdH QOs shown in Fig. 1 reveal well-resolved Landau levels at low magnetic fields. These QOs are expected to vanish above the QL magnetic field $B_{QL}$, when all carriers are confined in the lowest Landau level. This threshold $B_{QL}$ can be estimated from the last MR peak (Landau index $\nu = 1$) of the SdH QOs. For example, the QL field $B_{QL}$ of sample A4 shown in Fig. 2b is around 1.75 T, which is larger than the SdH oscillation frequency $F$ due to the phase shift ($F/B + \phi = \nu$)[39]. This characteristic field $B_{QL}$ can be influenced by the Zeeman effect. The relatively poor signal-to-noise ratio inherent to pulsed high magnetic field measurements in Figs. 2a and 2b obscures related details. Figure 2h shows the measured MR of sample A8 at static magnetic fields, in which the pronounced bulge at around 2 T signals the spin-splitting feature[40]. Figure 2i displays the extracted SdH QOs at different temperatures for magnetic field applied along the $b$-axis, clearly revealing the splitting peaks corresponding to the Landau band $\nu = 1-$ and $\nu = 1+$ (± denote different spins). For this sample, the threshold $B_{QL}$ can be identified by the observed MR peak of $\nu = 1-$ at around 2.25 T. According to the Zeeman splitting effect, the Landé g-factor can be estimated using $g \cdot m^*/2m_0 = F(1/B_+ - 1/B_-)$, where $m^*$ is the cyclotron effective mass and $m_0$ denotes the free-electron mass[41]. The SdH oscillation frequency $F$ is around 1.66 T as revealed by the fast Fourier transform (FFT) spectrum of these QOs shown in the inset of Fig. 2j. The cyclotron effective mass $m_b^*$ is estimated to be 0.015 $m_0$ by applying the Lifshitz–Kosevich formula to the temperature dependence of the oscillation amplitude (Fig. 2f). With these values, the calculated Landé g-factor is approximately 33, which is in good agreement with that obtained from anomalous Hall effect induced by Zeeman splitting[33]. Anomalous Hall effect was also observed in this work and the scaling behavior confirms the intrinsic Berry curvature as physical origin (Fig. S3) in this topological material[42-47]. Taken together, the MR results under a wide magnetic field range demonstrate the coexistence of both SdH and log-periodic QOs in a single sample and reveal a continuous field-driven transition from single-particle Landau quantization to interaction-driven DSI feature observed only when the system enters the QL regime.

**Angular-dependent QOs and the map of Fermi surface**

To fully characterize the SdH and log-periodic QOs, we further performed angular-dependent MR measurements of a typical sample (A1) under high magnetic fields (Fig. S4). The MR measured at 4.2 K (Fig. 3a) for various tilt angles within the *a-b* plane ($\theta_1$) exhibits well-defined SdH QOs in the low-field regime (periodic in $1/B$) (Fig. 3b), together with a transition to log-periodic QOs as the system enters the QL regime (Fig. 3c). MR data under different field orientations within the *a–c* plane ($\theta_2$) display similar behavior (Fig. 3d), with the oscillatory components after background subtraction identified as SdH and log-periodic QOs (Figs. 3e and 3f). The plot of Landau fan diagram (Fig. S5) for these SdH QOs yields frequencies $F_a$ = 10.87 T, $F_b$ = 0.99 T, and $F_c$ = 6.76 T for magnetic fields aligned with the crystallographic *a*, *b*, and *c* axes, respectively. The angular-dependent SdH oscillation frequencies (Fig. 3g) aligns well with the three-dimensional anisotropic Fermi surface model[30] $F^{3D} = F_b F_i / \sqrt{(F_b sin\theta)^2 + (F_i cos\theta)^2}$, where $\theta = \{\theta_2, \theta_1\}$ and $i = \{a, c\}$ (Fig. S6). The inset of Fig. 3g illustrates the schematic of the elliptical hole pocket, which has been seldom detected in earlier research. By analyzing the temperature-dependent SdH QOs, we extract the corresponding cyclotron masses $m_a^*$ = 0.132 $m_0$, $m_b^*$ = 0.015 $m_0$, and $m_c^*$ = 0.087 $m_0$, which are reproduced in multiple samples (Fig. 2j and Fig. S7). The cyclotron mass, denoted with a subscript indicating the direction of the magnetic field, represents the geometric mean of the effective mass (*m*) tensor components within the plane of electron motion perpendicular to the magnetic field. For an elliptical cross-section, this is expressed as $m_i^* = \sqrt{m_j m_k}$ with (i, j, k = $\{a, b, c\}$). Consequently, the band effective mass can be estimated as $m_a$ = 0.01 $m_0$, $m_b$ = 0.77 $m_0$, and $m_c$ = 0.02 $m_0$. These values reveal a dramatic anisotropy in the band curvature. The mapping of this hole pocket within the valence band offers crucial insights into the electronic structure of the intriguing topological material. Note that $F_a$ is larger than $F_c$ for the hole pocket, which is opposite to the feature of electron pocket[36]. Moreover, log-periodic QOs are difficult to be observed in electron-dominated HfTe$_5$ at low temperatures. Whether the band asymmetry plays a role for the different phenomena observed in hole-dominated and electron-dominated HfTe$_5$ needs future investigations.

The angular-dependent measurements also reveal the evolution of the log-periodic QOs under magnetic fields of different orientations. As shown in Fig. 3h, the scale factor $\lambda$ calculated from Figs. S8 to S10 varies with the tilt angle of magnetic field relative to *b* axis (Fig. 3h). In particular, at small angles, the log-periodic QOs follows a quasi-2D behavior, scaling with the perpendicular field component (Fig. 3i). At larger angles, the 2D approximation breaks down, revealing a 3D character consistent with the anisotropic Fermi surface deduced from SdH analysis. Aligning the magnetic field with the *a*- or *c*-axes results in a reduction of $\lambda$ (Fig. 3i). This observation is consistent with the theoretical framework for relativistic supercritical atomic quasi-bound states. The scale factor $\lambda$ follows the relation $\lambda = e^{\pi/s_0}$, where $s_0 = \sqrt{(Z^*\alpha^*)^2 - 1}$ is determined by two key physical parameters: the effective central charge $Z^*$ of the quasi-bound state and the effective fine-structure constant $\alpha^*$ of the material[11, 23, 24]. The central charge is expected to be fixed for this specific sample while the fine-structure constant $\alpha^* = e^2/4\pi\varepsilon\hbar v_F$ ($\varepsilon$ is dielectric constant) can vary due to the variance of the Fermi velocity, where. When the magnetic field is applied along the *b*-axis, cyclotron orbit is confined to the *a-c* plane. The significantly smaller band effective masses ($m_a$ = 0.01 $m_0$ and $m_c$ = 0.02 $m_0$) imply a lager Fermi velocity $v_F = \hbar k_F/m$ in this configuration. This aligns with previous spectroscopic findings for the sister compound ZrTe$_5$, where the Fermi velocity within the *a-c* plane is much larger compared to the other two[48]. When the magnetic field is oriented in-plane along the *a*- or *c*-axis, the cyclotron motion in the corresponding perpendicular planes (*b-c* or *a-b*) is largely influenced by the band curvature along the *b*-axis. The larger $m_b$ = 0.77 $m_0$ indicates relatively smaller Fermi velocities, consistent with previous reports[48]. The reduced $v_F$ leads to a larger effective fine-structure constant, thereby providing a plausible explanation for the reduction in the scale factor $\lambda$ of the log-periodic QOs under fields aligned along *a*- or *c*-axis. These results demonstrate that angular-dependent log-periodic QOs serve as a new and unique transport avenue for mapping the electronic landscape and Fermi velocity anisotropy in Dirac materials.

**Carrier dependence of QOs**

We summarize the MR curves of different samples measured at high fields and low temperature with the field along *b*-axis (Fig. S12), in which the low-field SdH QOs and high-field log-periodic QOs can be identified. Based on the experimental results of the series of density-tunable crystals, we obtain SdH oscillation frequency $F$, scale factor $\lambda$ of the log-periodic QOs (Fig. S11) and the onset magnetic field $B_c$ at which the first peak of log-periodic QOs appears. In Fig. 4a, the $\lambda$ and $B_c$ are plotted versus the SdH oscillation frequency $F$. It is found that $\lambda$ decreases and $B_c$ increases as $F$ becomes larger, indicating that both the log-periodic and SdH QOs are tunable via carrier density.

From these data, we explore the quantitative relationship between the two types of QOs. As discussed above, the scale factor $\lambda$ is mainly determined by the $Z^*$ and $\alpha^*$. For a given material in the same measurement configuration, the $\alpha^*$ is typically fixed, but the carrier density can be tuned to modify vacuum polarization, thereby changing $Z^*$. A higher carrier density leads to a larger effective central charge $Z^*$, resulting in an increased $s_0$ and thus reducing scale factor[15]. This decrease in $\lambda$ is attributed to weaker vacuum polarization. Using the relationship[15] $Z^* = Z/(1 + ZQln(1/(\kappa \cdot a)))$ where $\kappa$ is proportional to the Fermi wavevector $k_F$ ($\kappa = \beta k_F$), $a$ is a short-distance cutoff at the atomic scale and Q is a numerical factor of order 1. We have $\ln\lambda \propto 1/s_0 \propto 1/Z^* \propto \ln\left(\frac{1}{\kappa \cdot a}\right)$. This leads to $\lambda \propto k_F^{-1} \propto F^{-1/2} \propto n^{-1/3}$. Figure 4b shows $\lambda$ versus $F^{-1/2}$ for different samples with the fit following this theoretical formula. The strong agreement between experimental data and theory validates the quantitative relationship between the two types of QOs, which can be well explained by the carrier-density-tuned DSI quasi-bound states. Moreover, in our observation, for sample with a larger SdH oscillation frequency, the log-periodic QOs are less prominent and hard to be identified. A linear fit yields a critical field of $F$ = 2.71 T, which defines the upper limit for observing log-periodic QOs ($\lambda$ = 1). This threshold suggests a critical carrier density of around $2.52\times10^{16}$ cm$^{-3}$.

Furthermore, the carrier-density-dependent evolution of the SdH and log-periodic QOs enables us to plot both $B_c$ and $B_{QL}$ as functions of carrier density *n* (Fig. 4c). As shown in Fig. 4c, $B_c > B_{QL}$ in all samples, indicating that log-periodic QOs emerge only after all carriers occupy the lowest Landau level, when SdH QOs have already disappeared. Theoretically, the onset field $B_c$, marking the first peak of the log-periodic QOs, is closely linked to Thomas-Fermi screening[12, 23], which sets the maximum size of the quasi-bound states that retain DSI. A higher carrier density leads to a larger $B_c$ due to the decreased Thomas-Fermi screening length $\xi$. This length and the corresponding $B_c$ can be expressed as[11] $B_c = \frac{\hbar c}{s_0^2 e \xi^2}$ with $\xi^{-2} = \frac{4\pi e^2 dn}{d\mu}$, $\mu = \hbar v_F k_F$, and $n = k_F^3/3\pi^2 = \frac{1}{3\pi^2}\left(F\frac{2e}{\hbar}\right)^{\frac{3}{2}}$. Here $n$ is the hole carrier density from the Dirac band and $\mu$ is the chemical potential. We then have $B_c = \frac{AF^{0.5}}{\frac{C}{F+2DF\ln(BF^{-0.5})+D^2F(\ln(BF^{-0.5}))^2}-E}$, where $A = 2mec\sqrt{\frac{2e}{\hbar}}$, $B = \frac{1}{\beta a\sqrt{2e/\hbar}}$, $C = \frac{Z^2 e^3 m^2}{32\pi\varepsilon^2 \hbar^2}$, $D = ZQ$, $E = \hbar\pi$. Based on these formulas, the experimental data $B_c$ are well fitted (dotted line) as shown in Fig. 4c.

**Discussion**

The evolution from SdH to log-periodic QOs in HfTe$_5$ reflects the fundamental competition between kinetic and potential energy. The interplay between these two components, as the primary terms in the Hamiltonian, dictates the electronic behavior of quantum matter. At low magnetic fields, magnetotransport is dominated by the kinetic energy of itinerant carriers. Their cyclotron motions generate discrete Landau levels, producing SdH QOs periodic in $1/B$. As the field strength increases such that the system enters the QL regime, where all carriers occupy the lowest Landau level, the kinetic energy is effectively quenched. This suppression allows the potential energy to exert a dominant influence. In this regime, supercritical Coulomb interactions arising from charged impurities drive the emergence of discrete scale symmetry associated with atomic collapse[21, 22, 49-51]. Within this framework, the Thomas-Fermi screening length $\xi$ dictates the effective range of the Coulomb potential, defining the onset field $B_c$. In systems hosting relativistic Dirac fermions, once the magnetic length becomes shorter than $\xi$ and the Coulomb interaction strength exceeds the supercritical threshold, Dirac carriers and charged impurities form a series of quasi-bound states. The resulting energy spectrum obeys a discrete scaling law, $\varepsilon_{i+1}/\varepsilon_i = \lambda$, with the corresponding characteristic radii satisfying $R_i/R_{i+1} = \lambda$. Driven by resonant scattering of these quasi-bound states at the Fermi level, this geometric scaling manifests as log-periodic QOs in transport coefficients including the MR and Hall resistance. Crucially, our results unveil that vacuum polarization acts as a fundamental renormalization agent, modulating the effective impurity charge and thus tuning the scale factor of the DSI feature through carrier density modulation as we demonstrated in the experimental results.

In this work, the continuous evolution of QOs represents a fundamental crossover from single-particle Landau quantization to an interaction-driven energy landscape governed by relativistic QED. This transition delineates a physical regime shift that the low-field SdH QOs originate from the kinetic energy-dominated cyclotron motion of itinerant carriers and the high-field log-periodic QOs emerge when kinetic energy is quenched in the QL, allowing supercritical Coulomb potentials to dictate the energy landscape. By bridging these two distinct quantum regimes, our study establishes a unified framework where DSI supersedes Landau levels as the primary organizing principle of the energy spectrum (Fig. 4c) in the supercritical regime. Furthermore, vacuum polarization is revealed to serve as an important tuning knob that defines the stability and scaling of relativistic bound states in the solid-state system, which provides an analog for exploring QED effects.

Ultimately, our findings offer a unique perspective on the intricate interplay between single-particle quantization, few-body bound states and many-body electronic screening. The ability to modulate DSI via carrier density and QED many-body effect establishes a strategic pathway for realizing and manipulating scale-invariant quantum states across a diverse range of topological matter. These insights also guide the exploration of emergent novel symmetry in broader relativistic systems, such as van der Waals heterostructures and other Dirac materials.

**Methods**

**Sample synthesis**

Single crystals of HfTe$_5$ were synthesized using Te-flux method. In a typical procedure, Hf flakes and Te ingots with an atomic ratio of Hf:Te = 1:199 were loaded into a 5 mL alumina crucible. The crucible was then sealed within a quartz ampoule under vacuum. The assembly was heated to 800 °C and held at this temperature for 8-18 hours to homogenize the melt, followed by cooling to 600 °C over 12 hours. Subsequently, the temperature was slowly decreased to 475 °C at a controlled rate over 48 hours. After dwelling at 475 °C for 8-48 hours, the HfTe$_5$ crystals were separated from the residual tellurium flux by centrifugation of the quartz ampoule. The as-grown HfTe$_5$ crystals typically exhibited needle-like morphologies. The length is ranging from 0.2 to 3 mm. The width is ranging from 0.1 to 0.5 mm. The thickness is ranging from 0.01 and 0.3 mm.

**Characterization and transport measurements**

X-ray diffraction patterns were acquired utilizing a PANalytical Empyrean 2 diffractometer, which is equipped with a copper target that emits X-rays at a wavelength of 1.54 Å. Energy-dispersive spectroscopy (EDS) data were obtained utilizing ZEISS EVO MA 10 and BRUKER XFlash equipment. The transport experiments were conducted using a 14T-PPMS (Physical Property Measurement System, Quantum Design) and a pulsed high-magnetic-field environment at the Wuhan National Laboratory for High Magnetic Fields. An 8-nm palladium layer and a 50-nm gold layer were thermally evaporated onto the sample surface to form effective ohmic-contact electrodes. A 25-μm-diameter gold wire was subsequently attached to the Pd/Au electrodes with silver epoxy. The magnetoresistance signal was measured using the conventional four-probe technique, while the Hall effect and magnetoresistance signals were concurrently obtained by the six-probe approach.

**Data availability** Data measured or analyzed during this study are available from the corresponding author on reasonable request.

**Acknowledgements**

We acknowledge the support from the National Natural Science Foundation of China (No. 21BAA01133, 12374052, 12404215, 12488201), Guangdong Provincial Quantum Science Strategic Initiative (No. GDZX2401009, GDZX2401001), the National Key Research and Development Program of China (No. 2025YFA1411300), Quantum Science and Technology-National Science and Technology Major Project (2021ZD0302403), the Interdisciplinary program of Wuhan National High Magnetic Field Center (No. WHMFC2025017), Huazhong University of Science and Technology, Guangzhou Basic and Applied Basic Research Foundation (No. 2025A04J5405), Research Center for Magnetoelectric Physics of Guangdong Province (No. 2024B0303390001), Guangdong Provincial Key Laboratory of Magnetoelectric Physics and Devices (No. 2022B1212010008), and the Instrumental Analysis & Research Center, Sun Yat-sen University.

**Author contributions**

J.Y., N.T. Y.L. and J.L. contributed equally to this work. H.W. conceived and supervised the research. J.Y. and Y.L. grew the samples; N.T., J.Y., G.J., Y.L, J.L. and H.Z. carried out the experiments; H.W., Y.Liu, H.L. X.X. and J.W. proposed the theoretical model; H.W., J.Y., N.T., Y.Li, J.L., and Y.Liu analyzed the data; Z.W. and D.G. participated in the discussion; H.W., J.Y., N.T., J.L., Y.Liu and J.W. wrote and revised the manuscript with comments from all authors.

**Competing interests**

The authors declare no competing interests.

**Additional information**

**Supplementary information** is available for this paper.

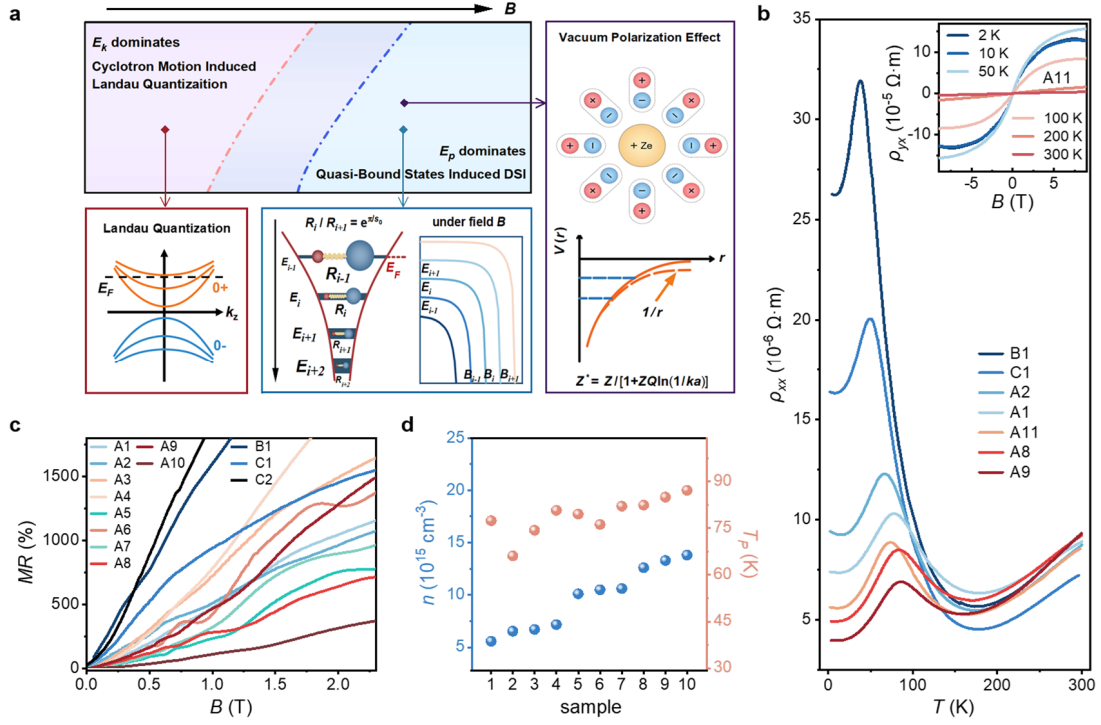

**Fig. 1 | Tunable magnetotransport responses via increasing magnetic field and varying carrier density. a,** Evolution from Landau quantization to discrete scale-invariant energy spectrum with increasing magnetic field. At low fields, the quantized cyclotron motion of carriers produces SdH QOs. In the quantum limit, Dirac carriers form discrete scale-invariant quasi-bound states with energies ($E_i$) and radii ($R_i$) following a geometric scaling, giving rise to log-periodic QOs. In this supercritical regime, the vacuum polarization effect can renormalize the effective charge and provide an avenue for tuning DSI. **b,** Temperature-dependent resistivity of HfTe$_5$ single crystals. Inset shows the Hall resistivity of a typical sample (A11), revealing hole-dominant transport at 2-300 K. **c,** MR of HfTe$_5$ single crystals in the field range of 0-2.5 T. **d,** Carrier density $n$ (left axis) and resistivity peak temperature $T_p$ (right axis) for samples. The carrier density is estimated based on the SdH QOs in (c) and $T_p$ is extracted from the temperature-dependent resistance curve. The Landau fan plot reveals frequencies $F$ ranging from 0.99 T to 1.81 T, giving $n \approx 5.57\times10^{15}$ to $1.38\times10^{16}$ cm$^{-3}$ across the samples. In low-density samples (such as B1, C1, C2), SdH QOs become undetectable, consistent with previous reports. The variance of resistivity, resistivity peak temperature and MR behavior demonstrate varying carrier density in this series of samples.

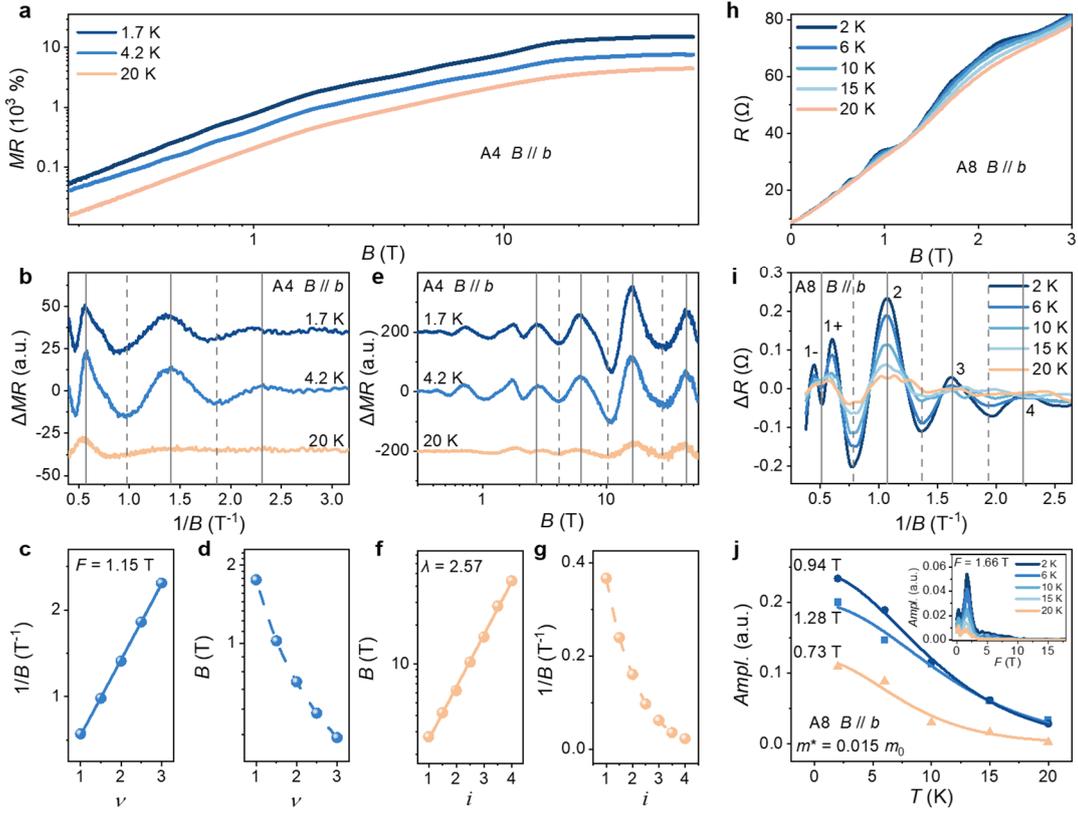

**Fig. 2 | Coexistence of SdH and log-periodic QOs. a,** MR behavior of sample A4 measured from 1.7 K to 20 K under pulsed magnetic fields (B // *b*-axis). The 4.2 K and 20 K curves are vertically offset for clarity. **b,** SdH QOs periodic in 1/*B* obtained by subtracting a smooth background. **c,** Landau fan diagram for the SdH QOs yields a frequency *F* = 1.15 T. **d,** Switch the vertical axis in **c** to logarithmic scale suppresses linear fitting, confirming the 1/*B*-periodicity. **e,** The log-periodic QOs in a semi-log plot. **f,** Semi-log plot of the fields corresponding to the oscillation maxima and minima in **e**. A scale factor *λ* = 2.57 is obtained from the linear fit. **g,** Switch the vertical axis in **f** to 1/*B* suppresses linear fitting, confirming the log-periodicity. **h,** MR behavior of sample A8 measured at static magnetic fields. **i,** The extracted SdH QOs show Zeeman splitting of the Landau band *ν* = 1 (± denote different spins). **j,** The temperature dependence of the SdH oscillatory amplitudes. The fits with the Lifshitz-Kosevich formula yield a cyclotron effective mass $m_b^* = 0.015\ m_0$, where $m_0$ is the free-electron mass. Inset is the FFT results of the SdH QOs revealing a frequency *F* = 1.66 T. The lower signal-to-noise ratio at low fields in pulsed field measurements (Figs. **2a-b**) leads to a reduced resolution of low-field SdH features compared to the steady-state data in Figs. **2h-j**.

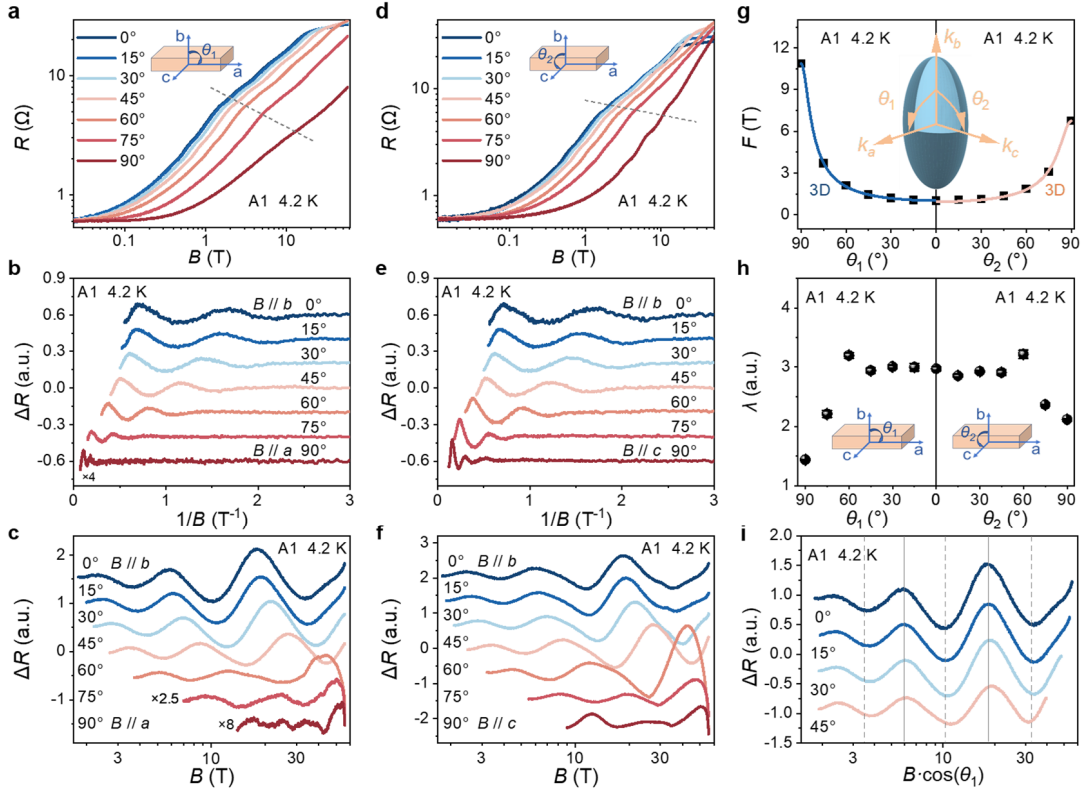

**Fig. 3 | Anisotropic Fermi surface of the hole pocket revealed by angular-dependent QOs. a,** MR of sample A1 for different magnetic field orientations in the *a-b* plane. **b,** Low-field SdH QOs after subtracting a smooth background from **a**. **c,** Extracted high-field log-periodic QOs rom **a**. **d,** MR of sample A1 for different magnetic field orientations in the *b-c* plane. **e-f,** SdH QOs (**e**) and log-periodic QOs (**f**) after removing background from **d**. **g,** Angular dependence of the SdH oscillation frequency, extracted from the linear fit of Landau fan diagram. The angular-dependent frequencies are well captured by a three-dimensional anisotropic Fermi surface model $F^{3D} = F_b F_i / \sqrt{(F_b \sin\theta)^2 + (F_i \cos\theta)^2}$, where $\theta = \{\theta_2, \theta_1\}$ and $i = \{a, c\}$. **h,** Angular dependence of scale factor $\lambda$ for the log-periodic QOs. The smaller $\lambda$ under in-plane fields aligned along *a*- or *c*-axes can be attributed to the reduced Fermi velocity in the *a-b* or *b-c* planes compared to *a-c* planes, which leads to a larger effective fine-structure $\alpha^* = e^2/4\pi\varepsilon\hbar v_F$ and thus a decreased $\lambda = e^{\pi/s_0}$ with $s_0 = \sqrt{(Z^*\alpha^*)^2 - 1}$. **i.** Log-periodic QOs in A1 as a function of the perpendicular component of magnetic field. At angles below 45°, the log-periodic features (indicated by dashed lines) exhibit a quasi-2D scaling. Note that at larger angles this approximation breaks down by the change of $\lambda$ as shown in panel **h**.

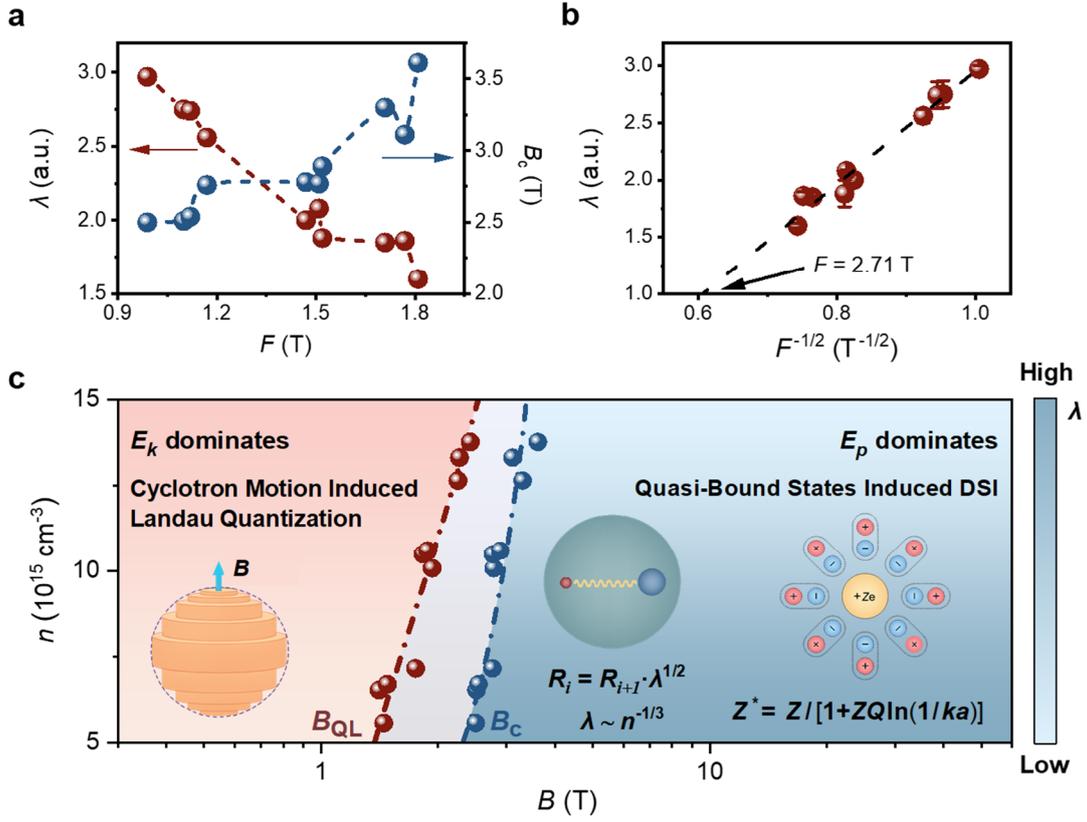

**Fig. 4 | The tunable QOs via carrier density modulation and magnetic field regime. a,** The scale factor $\lambda$ and the onset field $B_c$ of log-periodic QOs versus the SdH frequency $F$. **b,** The scale factor $\lambda$ as a function of $F^{-1/2}$. $\lambda$ can be quantitatively derived as $\lambda \propto k_F^{-1} \propto F^{-1/2} \propto n^{-1/3}$. A linear fit yields a critical field of $F = 2.71$ T, which defines the upper limit for observing log-periodic QOs. **c,** The transition from the SdH QOs characteristic of Landau quantization to the log-periodic QOs arising from scale-invariant energy structures. The $1/B$-periodic SdH QOs vanish in the QL regime, with the QL field $B_{QL}$ (red dots) identified by the final MR peak of the SdH QOs corresponding to Landau band $\nu = 1$. The onset field $B_c$ for log-periodic QOs (blue dots) consistently exceeds $B_{QL}$ and its dependence on carrier density is quantitatively described in the framework of supercritical quasi-bound atomic collapse states. The Thomas-Fermi screening length $\xi$ limits the effective range of the $1/r$ Coulomb potential, thereby defining the maximum radius of the quasi-bound states and the corresponding onset field $B_c$. The vacuum polarization renormalizes the effective central charge $Z^*$ and then directly modulate the scale factor $\lambda$.